\documentstyle[12pt,epsfig]{article}

\newcommand{\re}{\mathop{\mathrm{Re}}\nolimits}
\newcommand{\im}{\mathop{\mathrm{Im}}\nolimits}
\newcommand{\arsinh}{\mathop{\mathrm{arsinh}}\nolimits}
\newcommand{\arcosh}{\mathop{\mathrm{arcosh}}\nolimits}

\catcode`@=11
\newcount\@tempcntc
\def\@citex[#1]#2{\if@filesw\immediate\write\@auxout{\string\citation{#2}}\fi
  \@tempcnta\z@\@tempcntb\m@ne\def\@citea{}\@cite{\@for\@citeb:=#2\do
    {\@ifundefined
       {b@\@citeb}{\@citeo\@tempcntb\m@ne\@citea\def\@citea{,}{\bf
?}\@warning
       {Citation `\@citeb' on page \thepage \space undefined}}%
    {\setbox\z@\hbox{\global\@tempcntc0\csname b@\@citeb\endcsname\relax}%
     \ifnum\@tempcntc=\z@ \@citeo\@tempcntb\m@ne
       \@citea\def\@citea{,}\hbox{\csname b@\@citeb\endcsname}%
     \else
      \advance\@tempcntb\@ne
      \ifnum\@tempcntb=\@tempcntc
      \else\advance\@tempcntb\m@ne\@citeo
      \@tempcnta\@tempcntc\@tempcntb\@tempcntc\fi\fi}}\@citeo}{#1}}
\def\@citeo{\ifnum\@tempcnta>\@tempcntb\else\@citea\def\@citea{,}%
  \ifnum\@tempcnta=\@tempcntb\the\@tempcnta\else
   {\advance\@tempcnta\@ne\ifnum\@tempcnta=\@tempcntb \else
\def\@citea{--}\fi
    \advance\@tempcnta\m@ne\the\@tempcnta\@citea\the\@tempcntb}\fi\fi}
\catcode`@=12

\begin{document}
\title{\vskip-3cm{\baselineskip14pt
\centerline{\normalsize CERN-TH/2002-113\hfill ISSN 0418-9833}
\centerline{\normalsize DESY 02-073\hfill}
\centerline{\normalsize hep-ph/0205304\hfill}
\centerline{\normalsize May 2002\hfill}}
\vskip1.5cm
Elimination of Threshold Singularities in the Relation Between On-Shell and
Pole Widths}
\author{{\sc Bernd A. Kniehl,$^1$\thanks{Permanent address: II. Institut f\"ur
Theoretische Physik, Universit\"at Hamburg, Luruper Chaussee 149, 22761
Hamburg, Germany.} Caesar P. Palisoc,$^2$\thanks{Permanent address: National
Institute of Physics, University of the Philippines, Diliman, Quezon City
1101, Philippines.} Alberto Sirlin$^3$}\\
{\normalsize $^1$ CERN, Theoretical Physics Division, 1211 Geneva 23,
Switzerland}\\
{\normalsize $^2$ II. Institut f\"ur Theoretische Physik, Universit\"at
Hamburg,}\\
{\normalsize Luruper Chaussee 149, 22761 Hamburg, Germany}\\
{\normalsize $^3$ Department of Physics, New York University,}\\
{\normalsize 4 Washington Place, New York, New York 10003, USA}}

\date{}

\maketitle

\thispagestyle{empty}

\begin{abstract}
In a previous communication by two of us, Phys.\ Rev.\ Lett.\ {\bf81}, 1373
(1998), the gauge-dependent deviations of the on-shell mass and total decay
width from their gauge-independent pole counterparts were investigated at
leading order for the Higgs boson of the Standard Model.
In the case of the widths, the deviation was found to diverge at unphysical
thresholds, $m_H=2\sqrt{\xi_V}m_V$ ($V=W,Z$), in the $R_\xi$ gauge.
In this Brief Report, we demonstrate that these unphysical threshold
singularities are properly eliminated if a recently proposed definition of
wave-function renormalization for unstable particles is invoked.

\medskip

\noindent
PACS numbers: 11.15.Bt, 12.15.Lk, 14.80.Bn
\end{abstract}

\newpage

The unrenormalized propagator of a scalar boson, with four-momentum $q$, is of
the form
\begin{equation}
{\cal D}^{(u)}(s)=\frac{i}{s-M_0^2-A(s)},
\label{eq:unr}
\end{equation}
where $s=q^2$, $M_0$ is the bare mass, and $A(s)$ is the unrenormalized
self-energy.
In the case of the transverse propagator of a vector boson, there is an
additional factor $-(g^{\mu\nu}-q^\mu q^\nu/s)$ on the right-hand side of
Eq.~(\ref{eq:unr}).

In the conventional on-shell formulation, which most analyses in electroweak
perturbation theory are based on, the mass $M$ and total decay width $\Gamma$
of an unstable boson are defined as
\begin{eqnarray}
M^2&=&M_0^2+\re A(M^2),
\label{eq:mos}\\
M\Gamma&=&-\frac{\im A(M^2)}{1-\re A^\prime(M^2)},
\label{eq:gos}
\end{eqnarray}
respectively.
However, in gauge theories, Eqs.~(\ref{eq:mos}) and (\ref{eq:gos}) are known
to become gauge dependent at the next-to-next-to-leading order, {\it i.e.}, in
${\cal O}(g^4)$ and ${\cal O}(g^6)$, respectively, where $g$ is a generic
gauge coupling
\cite{sir,pa1,kn1,kn2}.
This problem can be solved by defining the mass and width in terms of the
complex-valued position of the propagator's pole,
\begin{equation}
\overline{s}=M_0^2+A(\overline{s}),
\label{eq:sba}
\end{equation}
which is gauge independent to all orders in perturbation theory
\cite{sir,wil,gam,gra}.
Fixing the pole mass $m_2$ and width $\Gamma_2$ through the parameterization
\cite{sir}
\begin{equation}
\overline{s}=m_2^2-im_2\Gamma_2,
\end{equation}
we have
\begin{eqnarray}
m_2^2&=&M_0^2+\re A(\overline{s}),
\label{eq:mpo}\\
m_2\Gamma_2&=&-\im A(\overline{s}).
\label{eq:gpo}
\end{eqnarray}
Alternative, gauge-independent definitions of mass and width based on
$\overline{s}$, with particular merits, were discussed in the literature
\cite{sir,wil}.
Recently, also gauge-independent definitions of partial decay widths that
properly add up to $\Gamma_2$ were introduced \cite{gra}.

Equation~(\ref{eq:mpo}) implies that the mass counterterm in the pole scheme
is given by $\delta m_2^2=m_2^2-M_0^2=\re A(\overline{s})$.
In order to complete the renormalization of Eq.~(\ref{eq:unr}), we also need
to specify an appropriate wave-function renormalization constant,
$Z=1-\delta Z$, so that the renormalized propagator,
\begin{eqnarray}
{\cal D}^{(r)}(s)&=&\frac{{\cal D}^{(u)}(s)}{Z}
\nonumber\\
&=&\frac{i}{s-m_2^2-{\cal S}^{(r)}(s)},
\label{eq:ren}
\end{eqnarray}
is ultraviolet (UV) finite.
An appropriate definition is \cite{pal}
\begin{equation}
Z=\frac{1}
{1+\left[\im A\left(\overline{s}\right)-\im A\left(m_2^2\right)\right]/
(m_2\Gamma_2)},
\label{eq:z}
\end{equation}
which allows us to rewrite Eq.~(\ref{eq:gpo}) as
\begin{equation}
m_2\Gamma_2=-Z\im A\left(m_2^2\right).
\label{eq:mg}
\end{equation}
In fact, $\delta m_2^2$ and $\delta Z$ thus defined are real and guarantee
that the renormalized self-energy,
\begin{equation}
{\cal S}^{(r)}(s)
=Z\left[A(s)-\delta m_2^2\right]+\delta Z\left(s-m_2^2\right),
\end{equation}
is UV finite to all orders \cite{pal,kn3,nek}.
Equation~(\ref{eq:z}) possesses a number of desirable properties.
On the one hand, it avoids threshold singularities that, in the conventional
on-shell scheme, appear in the radiatively corrected production and decay
rates of the Higgs boson as its mass approaches from below the pair-production
threshold of a vector boson \cite{pal}.
On the other hand, it precludes the occurrence of power-like infrared
divergences in the renormalized propagators of unstable particles that couple
to massless quanta, like the $W$ bosons and the quarks of the second and third
generations \cite{kn3,pa2}.
Finally, it allows one to systematically organize the order-by-order removal
of UV divergences in ${\cal S}^{(r)}(s)$ \cite{nek}.
In this Brief Report, we elaborate yet another virtue of Eq.~(\ref{eq:z}). 

Expanding Eqs.~(\ref{eq:mos}), (\ref{eq:gos}), (\ref{eq:mpo}), and
(\ref{eq:gpo}) about $s=m_2^2$ and combining the results, one obtains
\cite{kn1}
\begin{eqnarray}
\frac{M-m_2}{m_2}&=&-\frac{\Gamma_2}{2m_2}\im A^\prime\left(m_2^2\right)
+{\cal O}(g^6),
\label{eq:mdi}\\
\frac{\Gamma-\Gamma_2}{\Gamma_2}&=&
\im A^\prime\left(m_2^2\right)\left[\frac{\Gamma_2}{2m_2}
+\im A^\prime\left(m_2^2\right)\right]
-\frac{m_2\Gamma_2}{2}\im A^{\prime\prime}\left(m_2^2\right)+{\cal O}(g^6).
\label{eq:gdi}
\end{eqnarray}
In Ref.~\cite{kn1}, the gauge dependence of Eqs.~(\ref{eq:mdi}) and
(\ref{eq:gdi}) was analyzed for the Higgs boson in the Standard Model adopting
the $R_\xi$ gauge \cite{fuj}.
In the case of Eq.~(\ref{eq:gdi}), it was found that, for an arbitrary value
of $m_2$, unphysical threshold singularities, proportional to
$\left(m_2-2\sqrt{\xi_V}m_V\right)^{-1/2}$, occur as $\xi_V$ approaches from
below the point $m_2^2/\left(4m_V^2\right)$ ($V=W,Z$).
Here and in the following, $m_2$ and $\Gamma_2$ refer to the Higgs boson,
while $m_V$ denotes the pole mass of the intermediate boson $V$.
The purpose of this Brief Report is to demonstrate that the unphysical
threshold singularities encountered in Ref.~\cite{kn1} are eliminated if
Eq.~(\ref{eq:z}) is employed in a judicious manner.
For the time being, we disregard physical threshold singularities, which occur
independently of the choice of gauge if the Higgs-boson mass happens to have
the specific values $m_2=2m_V$ \cite{pal,bha} or $m_2=2m_f$, where $m_f$ is a
generic fermion mass, and we assume that the value of $m_2$ is sufficiently
far away from the points $2m_V$ and $2m_f$.
We shall return to the issue of physical threshold singularities in
Eqs.~(\ref{eq:mdi}) and (\ref{eq:gdi}) at the end of this Brief Report.

A one-loop expression for the unrenormalized Higgs-boson self-energy $A(s)$ in
the $R_\xi$ gauge may be found in Eq.~(8) of Ref.~\cite{pal}.
Detailed inspection reveals that the unphysical threshold singularity in
Eq.~(\ref{eq:gdi}) originates in $\im A^{\prime\prime}\left(m_2^2\right)$,
which contains the term $G_\mu m_2^2/\left(2\pi^2\sqrt2\right)
B_0^\prime\left(m_2^2,\xi_Vm_V^2,\xi_Vm_V^2\right)$, where $G_\mu$ is Fermi's
constant, $B_0$ is the scalar one-loop two-point integral in $D=4-2\epsilon$
space-time dimensions as given, {\it e.g.}, in Eq.~(9) of Ref.~\cite{pal}, and
the prime indicates differentiation with respect to the first argument.
In fact,
\begin{equation}
\im B_0^\prime\left(m_2^2,\xi_Vm_V^2,\xi_Vm_V^2\right)
=\frac{\pi a}{2m_2^2\sqrt{1-a}}\theta(1-a)+{\cal O}(\epsilon),
\label{eq:cul}
\end{equation}
where $a=4\xi_Vm_V^2/m_2^2$, exhibits the type of singularity mentioned above.

We now illustrate how this singularity is eliminated by consistently working
in the pole scheme \cite{pal,kn3}.
We start by observing that Eq.~(\ref{eq:gdi}) is based on the expansion
\cite{kn1}
\begin{eqnarray}
m_2\Gamma_2&=&-\im A\left(m_2^2\right)\left\{1+\re A^\prime\left(m_2^2\right)
+\left[\re A^\prime\left(m_2^2\right)\right]^2
-\frac{1}{2}\im A\left(m_2^2\right)\im A^{\prime\prime}\left(m_2^2\right)
\right.
\nonumber\\
&&{}+\left.{\cal O}(g^6)\right\}.
\label{eq:exp}
\end{eqnarray}
Here, it is tacitly assumed that $A(s)$ is analytic near $s=m_2^2$, so that
the Taylor expansion can be performed.
In most cases, this assumption is valid.
However, $A(s)$ possesses a branch point if $s$ is at a threshold.
As a consequence, at a given two-particle threshold $m_2=m_A+m_B$, the
derivatives $A^{(n)}\left(m_2^2\right)$ ($n=1,2,\ldots$) develop threshold
singularities proportional to $|m_2-m_A-m_B|^{-1/2}$ or worse.
The latter appear in $\re A^{(n)}\left(m_2^2\right)$ 
[$\im A^{(n)}\left(m_2^2\right)$] as $m_2$ approaches the threshold from below
(above).
In the case of the Higgs boson, the problems start at $n=1$ for $m_2=2m_V$
\cite{pal,bha} and at $n=2$ for the residual two-particle thresholds,
$m_2=2\xi_Vm_V$ \cite{kn1} and $m_2=2m_f$.
The solutions to all these problems emerge by undoing the Taylor expansions.
In Ref.~\cite{pal}, this was illustrated for $\re A^\prime\left(m_2^2\right)$
at $m_2=2m_V$.
Here, we consider $\im A^{\prime\prime}\left(m_2^2\right)$ at $m_2=2\xi_Vm_V$,
which is relevant for the investigation of the gauge dependence of
Eq.~(\ref{eq:gdi}) \cite{kn1}.

Inserting Eq.~(\ref{eq:z}) in Eq.~(\ref{eq:mg}) and expanding in powers of
$\left[\im A(\overline{s})-\im A\left(m_2^2\right)\right]/(m_2\Gamma_2)$, we
obtain
\begin{equation}
m_2\Gamma_2=-\im A\left(m_2^2\right)\left\{
1-\frac{\im A(\overline{s})-\im A\left(m_2^2\right)}{m_2\Gamma_2}
+\left[\frac{\im A(\overline{s})-\im A\left(m_2^2\right)}{m_2\Gamma_2}
\right]^2+{\cal O}(g^6)\right\}.
\label{eq:exz}
\end{equation}
It is important to note that
$\left[\im A(\overline{s})-\im A\left(m_2^2\right)\right]/(m_2\Gamma_2)$
involves a finite difference, rather than a derivative.
Due to this fact, and as we shall explicitly show later, it is free from
threshold singularities.
Comparison of Eqs.~(\ref{eq:exp}) and (\ref{eq:exz}) shows then that the
threshold singularities emerging from $\im A^{\prime\prime}\left(m_2^2\right)$
are avoided if this amplitude is replaced according to the substitution rule
\begin{equation}
\im A^{\prime\prime}\left(m_2^2\right)=-\frac{2}{m_2\Gamma_2}
\left[\frac{\im A(\overline{s})-\im A\left(m_2^2\right)}{m_2\Gamma_2}
+\re A^\prime\left(m_2^2\right)\right]+{\cal O}(g^4).
\label{eq:app}
\end{equation}
Away from thresholds, the expansion of $A(\overline{s})$ about $m_2^2$ is
valid, and both sides of Eq.~(\ref{eq:app}) are well defined.
However, at the unphysical threshold $m_2=2\xi_Vm_V$, such an expansion breaks
down, $\im A^{\prime\prime}\left(m_2^2\right)$ diverges, and only the
right-hand side of Eq.~(\ref{eq:app}) remains well defined.
The substitution in Eq.~(\ref{eq:app}) is equivalent to replacing
Eq.~(\ref{eq:cul}) by
\begin{equation}
\im B_0^\prime\left(m_2^2,\xi_Vm_V^2,\xi_Vm_V^2\right)
=\frac{\re B_0\left(\overline{s},\xi_Vm_V^2,\xi_Vm_V^2\right)
-\re B_0\left(m_2^2,\xi_Vm_V^2,\xi_Vm_V^2\right)}{m_2\Gamma_2}+{\cal O}(g^2).
\label{eq:sub}
\end{equation}
Using the expression for $B_0\left(s,m_V^2,m_V^2\right)$ given in Eq.~(13) of
Ref.~\cite{pal} and introducing the auxiliary function
\begin{equation}
f(z)=-2\sqrt{1-z}\arsinh\sqrt{-\frac{1}{z}},
\end{equation}
we can rewrite Eq.~(\ref{eq:sub}) as
\begin{equation}
\im B_0^\prime\left(m_2^2,\xi_Vm_V^2,\xi_Vm_V^2\right)
=\frac{\re f\left(4\xi_Vm_V^2/\overline{s}\right)-\re f(a-i\varepsilon)}
{m_2\Gamma_2}
+{\cal O}(\epsilon)+{\cal O}(g^2),
\label{eq:su1}
\end{equation}
where $a$ is defined below Eq.~(\ref{eq:cul}).
We have
\begin{eqnarray}
\re f\left(\frac{4\xi_Vm_V^2}{\overline{s}}\right)
&=&-\frac{\sqrt2}{b}\left\{\frac{1}{2}\sqrt{b(c+b)-a}\right.
\nonumber\\
&&{}\times\ln\left[\frac{1}{a}\left(b+c+\sqrt{(b-1)(c+a-1)}
+\sqrt{(b+1)(c-a+1)}\right)\right]
\nonumber\\
&&{}+\left.\sqrt{b(c-b)+a}
\arctan\frac{\sqrt{b+1}+\sqrt{c-a+1}}{\sqrt{b-1}+\sqrt{c+a-1}}\right\},
\\
\re f(a-i\varepsilon)&=&-2\sqrt{1-a}\arcosh\sqrt\frac{1}{a}\theta(1-a)
-2\sqrt{a-1}\arcsin\sqrt\frac{1}{a}\theta(a-1),
\end{eqnarray}
where $b=\sqrt{1+\gamma^2}$ and $c=\sqrt{(a-1)^2+\gamma^2}$, with
$\gamma=\Gamma_2/m_2$.
We note that the discontinuity
$f(a+i\varepsilon)-f(a-i\varepsilon)=-2\pi i\sqrt{1-a}\theta(1-a)$ is purely
imaginary, so that $\re f(a+i\varepsilon)=\re f(a-i\varepsilon)$.
At $m_2=2\sqrt{\xi_V}m_V$, Eq.~(\ref{eq:su1}) becomes
\begin{equation}
\im B_0^\prime\left(m_2^2,\xi_Vm_V^2,\xi_Vm_V^2\right)
=-\frac{1}{m_2^2}\left[\frac{\pi}{\sqrt{2\gamma}}\left(1-\frac{\gamma}{2}
-\frac{3}{8}\gamma^2\right)+\frac{4}{3}\gamma
+{\cal O}\left(\gamma^{5/2}\right)\right]+{\cal O}(\epsilon)+O(g^2),
\end{equation}
{\it i.e.}, the unphysical threshold singularity is automatically regularized
in the pole scheme by the width $\Gamma_2$ of the primary particle.

\begin{figure}[ht]
\begin{center}
\centerline{\epsfig{figure=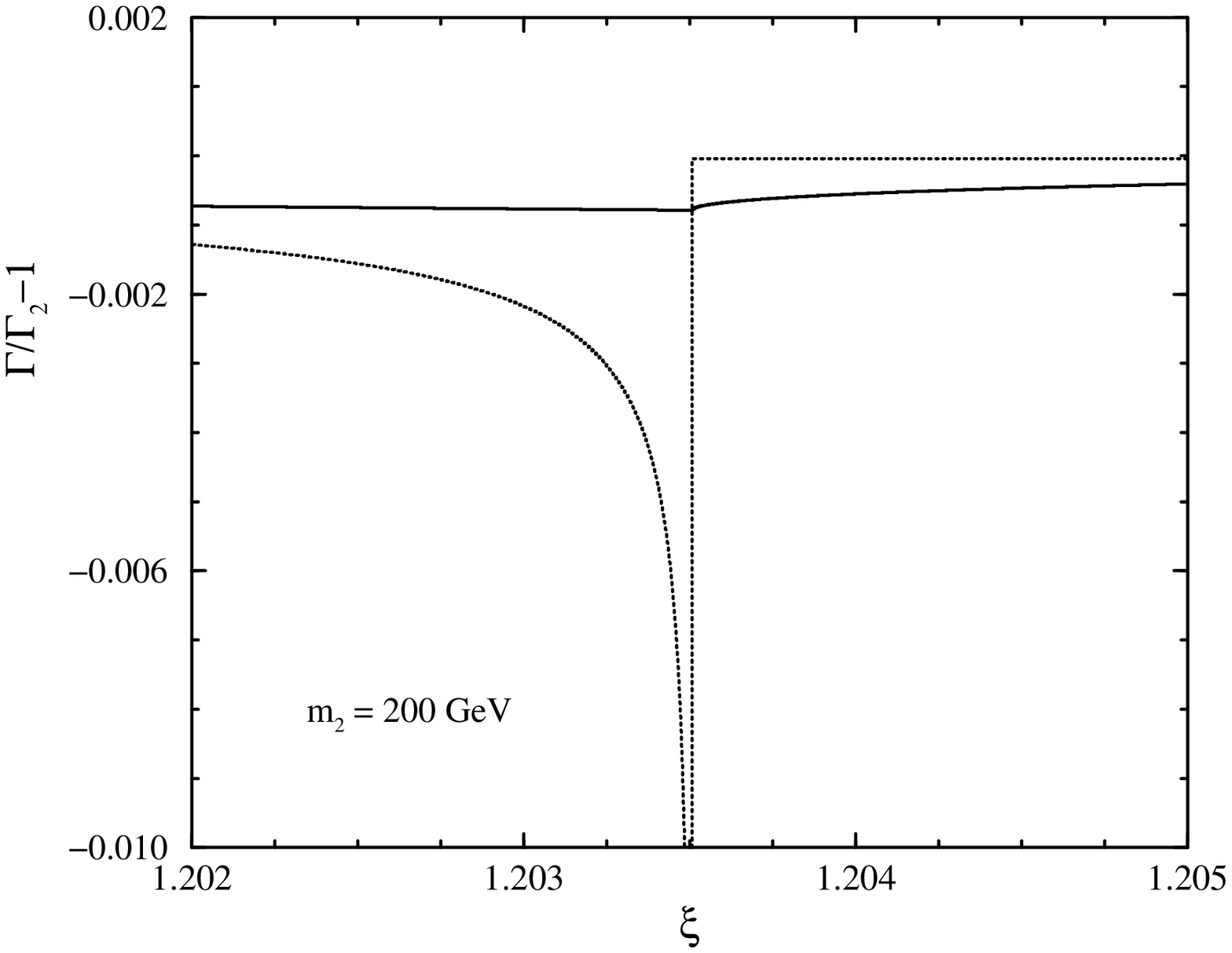,width=16cm}}
\caption{Relative deviation of the on-shell width $\Gamma$ from the pole
width $\Gamma_2$ for a Higgs boson with pole mass $m_2=200$~GeV as a function
of the gauge parameter $\xi$ in the vicinity of the point $m_2^2/(4m_Z^2)$.
The unphysical threshold singularity originating in Eq.~(\ref{eq:cul}) (dotted
line) is eliminated by applying the substitution rule of Eq.~(\ref{eq:su1})
(solid line), which is a consequence of invoking Eq.~(\ref{eq:z}).}
\label{fig:one}
\end{center}
\end{figure}

In our numerical analysis, we use the pole-mass values $m_W=80.391$~GeV and
$m_Z=91.154$~GeV, which are extracted from the measured values \cite{pdg} as
described in Ref.~\cite{pal}, and adopt the residual input parameters from
Ref.~\cite{pdg}.
For definiteness, we choose $m_2=200$~GeV and evaluate $\Gamma_2$ in the Born
approximation.
For simplicity, we set $\xi=\xi_W=\xi_Z$.
In Fig.~\ref{fig:one}, we show the $\xi$ dependence of
$(\Gamma-\Gamma_2)/\Gamma_2$ given by Eq.~(\ref{eq:gdi}) in the vicinity of
the point $m_2^2/\left(4m_Z^2\right)$.
The dotted and solid lines are evaluated using Eqs.~(\ref{eq:cul}) and
(\ref{eq:su1}), respectively.
The dotted line corresponds to the solid line in Fig.~1(a) of Ref.~\cite{kn1}
and exhibits the familiar abyss at $\xi=m_2^2/\left(4m_Z^2\right)$.
Obviously, this unphysical threshold singularity is absent in the solid line,
which smoothly interpolates across the threshold region and merges with the
dotted line sufficiently far away from the threshold.
A similar discussion applies to the second abyss, at
$\xi=m_2^2/\left(4m_W^2\right)$, which is not visible in Fig.~\ref{fig:one}. 

Finally, we return to the physical threshold singularities.
From the discussion below Eq.~(\ref{eq:exp}) it follows that threshold
singularities analogous to the one displayed in Eq.~(\ref{eq:cul}) also affect
Eq.~(\ref{eq:mdi}) for $m_2=2m_V$ and Eq.~(\ref{eq:gdi}) for $m_2=2m_V$ and
$m_2=2m_f$.
They may be eliminated in a very similar way, by applying the substitution
rule of Eq.~(\ref{eq:su1}), with $\xi_Vm_V^2$ replaced by $m_V^2$ or $m_f^2$.

\smallskip

B.A.K. and C.P.P. are grateful to the CERN Theoretical Physics Division and
the Second Institute for Theoretical Physics of Hamburg University,
respectively, for their hospitality during visits when this Brief Report was
finalized.
The work of B.A.K. was supported in part by the Deutsche
Forschungsgemeinschaft through Grant No.\ KN~365/1-1 and by the
Bundesministerium f\"ur Bildung und Forschung through Grant No.\
\break 05~HT1GUA/4.
The work of C.P.P. was supported by the Office of the Vice President for
Academic Affairs of the University of the Philippines.
The work of A.S. was supported in part by the Alexander von Humboldt
Foundation through Research Award No.\ IV~USA~1051120~USS and by the
National Science Foundation through Grant No.\ PHY-0070787.

\newpage

\end{document}